\documentclass[12pt]{article}
\usepackage{authblk}
\usepackage{times}
\usepackage{helvet}
\usepackage{courier}
\usepackage{graphicx}
\usepackage{caption}
\usepackage{subcaption}
\providecommand{\keywords}[1]{\textbf{\textit{Keywords---}}#1}

\title{Not All Explanations are Created Equal: Investigating the Pitfalls of Current XAI Evaluation}
\author{Joe Shymanski}
\author{Jacob Brue}
\author{Sandip Sen}
\affil{{\small
    Tandy School of Computer Science\\
    The University of Tulsa\\
    \{joe-shymanski, jacob-brue, sandip-sen\}@utulsa.edu\\
}}
\date{} 

\begin{document}

\maketitle

\begin{abstract}
Explainable Artificial Intelligence (XAI) aims to create transparency in modern AI models by offering explanations of the models to human users. There are many ways in which researchers have attempted to evaluate the quality of these XAI models, such as user studies or proposed objective metrics like “fidelity”. However, these current XAI evaluation techniques are \textit{ad hoc} at best and not generalizable. Thus, most studies done within this field conduct simple user surveys to analyze the difference between no explanations and those generated by their proposed solution. We do not find this to provide adequate evidence that the explanations generated are of good quality since we believe any kind of explanation will be “better” in most metrics when compared to none at all. Thus, our study looks to highlight this pitfall: most explanations, regardless of quality or correctness, will increase user satisfaction. We also propose that emphasis should be placed on {\em actionable explanations}. We demonstrate the validity of both of our claims using an agent assistant to teach chess concepts to users. The results of this chapter will act as a call to action in the field of XAI for more comprehensive evaluation techniques for future research in order to prove explanation quality beyond user satisfaction. Additionally, we present an analysis of the scenarios in which placebic or actionable explanations would be most useful.

\end{abstract}

\section{Introduction}
\label{sec:introduction}

The field of machine learning has been experiencing unprecedented growth over the past decade \cite{aggarwal2022has}. Often, this growth is fueled by major innovations in deep learning technology. Machine learning models have proven that function approximators can discover deep patterns within data. 
While modern machine learning models have been very successful at finding solutions for learning tasks, many challenges remain for translating learned knowledge from trained models to human understanding. These challenges span ethical, scientific, and practical goals, and are often critical barriers to the adoption of machine learning models 
\cite{langer2021we}. Some of these challenges will be elaborated in section \ref{sec:related-work}, such as model validation, privacy, and user satisfaction.
These challenges apply across many machine learning applications, so solutions are crucial for the continued adoption of machine learning. The field of explainable artificial intelligence (XAI) was created for research that addresses these challenges directly through the generation of explanations 
\cite{langer2021we, renftle2022explaining}.

Since XAI systems facilitate human-AI interaction, explanation generation has become a critical piece of the process and infrastructure for the effective deployment of machine learning models. While the requirements and restrictions for an explanation generation algorithm are diverse, the underlying challenge of human engagement is shared. As a result, human-agent interaction can provide valuable insights into the challenges and solutions of developing an XAI algorithm \cite{rosenfeld2019explainability}. 
It is appropriate to design the explanation generation algorithms as an AI agent. Any explanation generation algorithm perceives a model and chooses between different communication actions to explain the algorithm to a user. This allows explanation generation algorithms to be studied through the field of intelligent agents. Explanatory agents solve problems inherent to opaque machine learning models by offering transparency to build trust, satisfaction, and understanding for users \cite{lavender2023relative}.

Even though many explanation generation algorithms have been proposed and used, what makes an explanation good is still fuzzy and underdefined. 
Some have looked to studying human explanations in the social sciences as a model for good explanations
\cite{miller2019explanation, mohseni2021quantitative}. Several characteristics 
of human explanations have been identified as potentially important for XAI. Human explanations are generally conversational. They are contrastive, and they rely on shared biases.
An alternative view of what makes an explanation good is as a measure of how well or how often it achieves the set of goals for which it was designed. These goals include effective teaching, increasing satisfaction and trust, highlighting ethical/privacy concerns, highlighting model flaws, and discovering novel insights. We will focus primarily on the goal of increasing user satisfaction as well as user understanding of the model. These challenges are the most prevalent for agents designed to interface primarily with non-domain experts \cite{mohseni2021multidisciplinary}. Evaluation techniques for determining fidelity or robustness are beyond the scope of this study.

Evaluating the explanations of models is its own challenge. 
The variety of goals of XAI systems is large, and the variety of evaluation techniques that have been used to measure the effectiveness of explanations is even larger
\cite{mohseni2021multidisciplinary, chromik2020taxonomy, keane2021if, belaid2022we, rosenfeld2021better, mohseni2018human}. There is no standard for how XAI should be evaluated. The most common evaluation methods involve user studies. 
We have found throughout the literature that XAI evaluation techniques are often inadequate. While a multitude of measures have been proposed for calculating objective metrics such as fidelity, user studies are a very common approach for the more subjective human factors like satisfaction. Many times, researchers will compare a group receiving explanations to a control group with no explanations whatsoever. We believe this technique to be flawed in nature and seek to prove that most explanations, outside of obviously false statements, will boost user satisfaction when compared to a group with no explanations
\cite{eiband2019impact, ehsan2021explainability, nourani2019effects}. This effect has been labelled ``placebic''. Placebic explanations have seen some initial research that suggests they can produce similar satisfaction for users to more informative explanations. 
In addition, we also seek to propose a better baseline against which the output of future XAI models can be compared, rather than no explanation. 

We present our early analysis, initial study, and current results. This research is still in progress, and we will highlight the steps we have left after our conclusions.
In the introduction, we have discussed the development of the field of XAI and its purpose, how it is evaluated, and a recommendation to change a flawed pattern within those evaluations. In section \ref{sec:research-goals}, we will highlight the goals of our research study. In section \ref{sec:related-work}, we discuss related work surrounding our research, including analysis of the goals of XAI, findings on explanations within the social sciences, current methods for XAI evaluation, and evidence for the effects of placebic explanations. In section \ref{sec:domain-analysis}, we analyze the domains in which certain explanation types are most effective. In section \ref{sec:research-hypotheses}, we state our hypothesis for the study. In section \ref{sec:methodology}, we describe our methodology for the study, tracing the steps through our experiments. In section \ref{sec:results-and-data-analysis}, we report the results of our experiments. In section \ref{sec:discussion}, we discuss the findings from the data we gathered. In section \ref{sec:conclusions}, we share our conclusions. In section \ref{sec:future-work}, we discuss the remaining steps in our research, as well as present future work that can be undertaken to extend our research across other domains, performance metrics, and study designs.




\section{Research Goals}
\label{sec:research-goals}

This study aims to demonstrate that user satisfaction with an explanation is more favorable compared to user satisfaction without an explanation, regardless of the quality of the explanation. We will do this by using placebic explanations. This expectation is based on the principle that users positively receive an agent that communicates frequently over a less communicative agent, regardless of the content of the language. Another possible pathway to arrive at this result is the social concept that a willingness to discuss a topic lends credibility to a conversational agent without requiring logical arguments.

\section{Related Work}
\label{sec:related-work}

Langer et al.\ identified five classes of stakeholders for Machine Learning models \cite{langer2021we}. These include users, system developers, affected parties, deployers, and regulators. Each class has several desiderata from XAI, sometimes overlapping among the classes.
One desiderata in model development that benefits from the increased understandability of XAI is model validation. There are a plethora of reasons why a model can appear to succeed in a test scenario, but fail in its deployment. 
XAI can increase the likelihood for a model inaccuracy to be caught before deployment and allows for expert correction.
Another desiderata comes from privacy and ethical concerns. Many systems work with or around sensitive data. XAI can be used to find and address inequality, biases, and privacy risks present within a model.
Some models learn a novel solution to a learning task. Often the solution of the model contains scientific discoveries within the domain of the task. XAI increases the ability to translate new insights from the model. 
Even a solution that isn't scientifically novel can provide new learning to a person unfamiliar with a task. A unique challenge for explainable agents is to teach new insights to model users.
Finally, models on their own do not generally inspire trust or satisfaction from users. Whether it is the primary goal, or more often a secondary goal, user satisfaction and trust are more likely to result from understandable models. 

Miller argues that good explanations in XAI should be similar to those produced in human conversation \cite{miller2019explanation}. He reviews the literature from the social sciences to describe the factors of human explanations. 
Mohseni et al.\ discuss the divide between objective computational methods for evaluating XAI fidelity and the study of human factors like satisfaction and trust \cite{mohseni2018human}.

Chromik and Schuessler survey human subject XAI evaluation methods in order to produce a taxonomy \cite{chromik2020taxonomy}. Our study fits within the taxonomy. Mohseni and Zarei conducted a survey of XAI literature to aggregate and classify the different XAI design goals and the different methods of evaluating explainable AI systems \cite{mohseni2021multidisciplinary}. They identified 5 main categories of evaluation measures, including mental models, usefulness and satisfaction, user trust and reliance, human-AI task performance, and computational measures. Our study focuses on the second and fourth evaluation measures (satisfaction and human-AI task performance), which represent the most common evaluation measures used in the XAI field according to the study.

Belaid et al.\ designed a computational benchmark for XAI evaluation \cite{belaid2022we}. It is a composite of multiple tests to help score model fidelity.
Rosenfeld argues that XAI evaluation should be done primarily through objective computed metrics \cite{rosenfeld2021better}.
Keane et al.\ found that user studies were underrepresented as an evaluation metric within counterfactual-based XAI methods \cite{keane2021if}. They claim that evaluations that neglect user-centered evaluation may fail to prove that the algorithm achieves its goals for human users. 

Eiband et al.\ studied the effect of placebic explanations on perceived trust within a prototype nutrition recommendation application \cite{eiband2019impact}. They found that placebic explanations promote an increase in perceived trust and understandability similar to ``real explanations''. They found this pattern within a small study size. 
Ehsan and Riedl highlight placebic explanations as an explainability pitfall, an unanticipated behavior of an XAI algorithm that can have negative consequences \cite{ehsan2021explainability}. 
The study by Nourani et al.\ found that explanations that go against a user's expectations can cause a decrease in their perceived model accuracy \cite{nourani2019effects}. This effect is different from the one we are studying because our placebic explanations do not go against user expectations.

\section{Domain Analysis}
\label{sec:domain-analysis}

\begin{figure}
    \centering
    \begin{subfigure}{.69\textwidth}
        \centering
        \includegraphics[width=\textwidth]{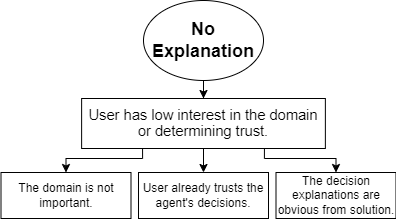}
        \caption{No explanations.}
        \label{fig:none-diagram}
    \end{subfigure}
    \begin{subfigure}{.69\textwidth}
        \centering
        \includegraphics[width=\textwidth]{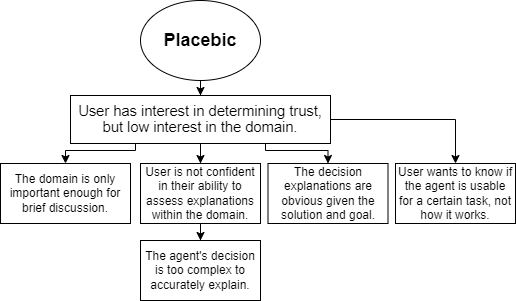}
        \caption{Placebic explanations.}
        \label{fig:placebic-diagram}
    \end{subfigure}
    \begin{subfigure}{.69\textwidth}
        \centering
        \includegraphics[width=\textwidth]{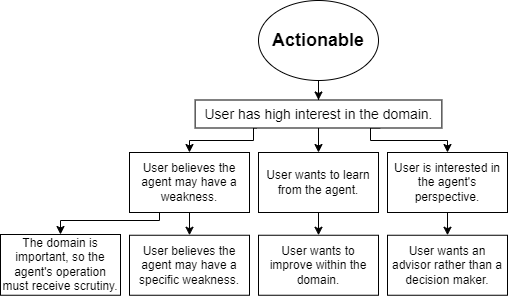}
        \caption{Actionable explanations.}
        \label{fig:actionable-diagram}
    \end{subfigure}
    \caption{Situational preferences for each explanation type.}
    \label{fig:exp-diagram}
\end{figure}

Actionable explanations are not necessarily ideal in every scenario. Figure~\ref{fig:exp-diagram}
demonstrates how the user's preferences and the problem domain can influence the desired explanation type. Note that these are not mutually exclusive categories, and any given situation may apply to several explanation types at once. The bidirectionality present within these human-agent systems requires the agent to understand the human's intentions and the human to properly express them. Relative weighting could then be used to determine the proper explanation. An iterative process should be used if there is little user information and the problem is not time sensitive like in real-time XAI systems.

We believe that users who are either not engaging with the system or do not need to gain further trust in the agent do not require any explanations whatsoever. These users would be wasting their time and energy when presented with the explanations given by the model. A self-driving car that has taken the same detour each day to avoid construction need not explain itself every time; the user trusts the car's decision by this point.

Placebic explanations are suitable for scenarios where the user desires a sanity check from the agent. They are also suited well for scenarios where the user is not an expert in the problem and does not need verbose, domain-specific rhetoric in the explanation. This makes them appropriate as baselines for evaluation. A patient may want to know that their software-recommended prescription is used to treat their medical issue, but they probably do not want to know the chemical processes it initiates.

There are a number of scenarios in which actionable explanations are most useful. Any highly important problem that the agent will attempt to solve on its own should warrant actionable explanations to the user overseeing its efforts. If such an important problem is to be solved ultimately by a user, such detailed explanations are required so the human can make informed decisions.

Additionally, users who are interested in learning a strategy to solve a problem on their own successfully would be interested in receiving actionable details. The experiments within this chapter focus on this exact situation. Hence we expect actionable explanations to be more effective compared to placebic explanations in our domain.

\section{Research Hypotheses}
\label{sec:research-hypotheses}

A coherent explanation can significantly increase an engaged user’s satisfaction over no explanation, even if the explanation is ineffective at increasing demonstrable improvement in model understanding.
Thus, we think placebic explanations will cause higher satisfaction in users than no explanations. Additionally, these placebic explanations will be just as satisfying for users as higher quality ``actionable'' explanations. An actionable explanation is one which, unlike a placebic explanation, offers relevant and potentially new information to the user which can help them decide on the best action to take. Actionable explanations aim to provide reasoning that the human recipient can utilize in future scenarios on their own. Thus, we would expect to see these actionable explanations boost user task performance. We might also expect the users to recognize the increased explanatory power of these high-quality explanations since they are presumably more helpful for future incentive-based tasks. Thus, we develop 5 hypotheses for this study:

\begin{description}
    \item[H1:] Placebic explanations have higher user satisfaction ratings than no explanations.
    \item[H2:] Placebic explanations do not have significantly different satisfaction ratings than actionable explanations.
    \item[H3:] Actionable explanations increase user understanding via task performance more successfully than no explanations.
    \item[H4:] Actionable explanations increase user understanding via task performance more successfully than placebic explanations.
    \item[H5:] Users perceive actionable explanations to have higher explanatory power than placebic explanations.
\end{description}

\section{Methodology}
\label{sec:methodology}

For our experimentation, we chose the domain of chess and, in particular, the effective attacking concepts of {\em forks} and {\em pins}. The best model moves were recommended to the users (learners), and explanations were provided to supplement and teach concepts during a practice section. Then, user comprehension was tested through a series of test scenarios. We evaluated three conditions or protocols corresponding to explanation types, or lack thereof, associated with agent-recommended moves: (i) no explanations, (ii) placebic explanations, and (iii) actionable explanations. For the purposes of testing and consistency across users, the explanations were human-generated by the researchers who were more than experienced with XAI explanations and with the tactics presented in each puzzle.

\subsection{Domain}
\label{sec:domain}

This study had the user complete several chess puzzles illustrating a couple of key chess concepts. A single puzzle involved a given board state where one sequence of moves would result in the most favored position for the user, while any other set of moves would result in significantly less favorable positions, i.e., there was only one correct move choice sequence that was the ``answer'' to the puzzle. All of the puzzles were curated to be relatively basic and only contain the tactic specified. They all featured exactly two moves for the user to make correctly, with a single computer move made for the black pieces in between. The subjects always played as the white pieces, and everybody received the same puzzles in the same order, no matter the explanation protocol.

\subsection{Agent Interface}
\label{sec:agent-interface}

\begin{figure}
    \centering
    \begin{subfigure}{\columnwidth}
        \centering
        \includegraphics[width=\textwidth]{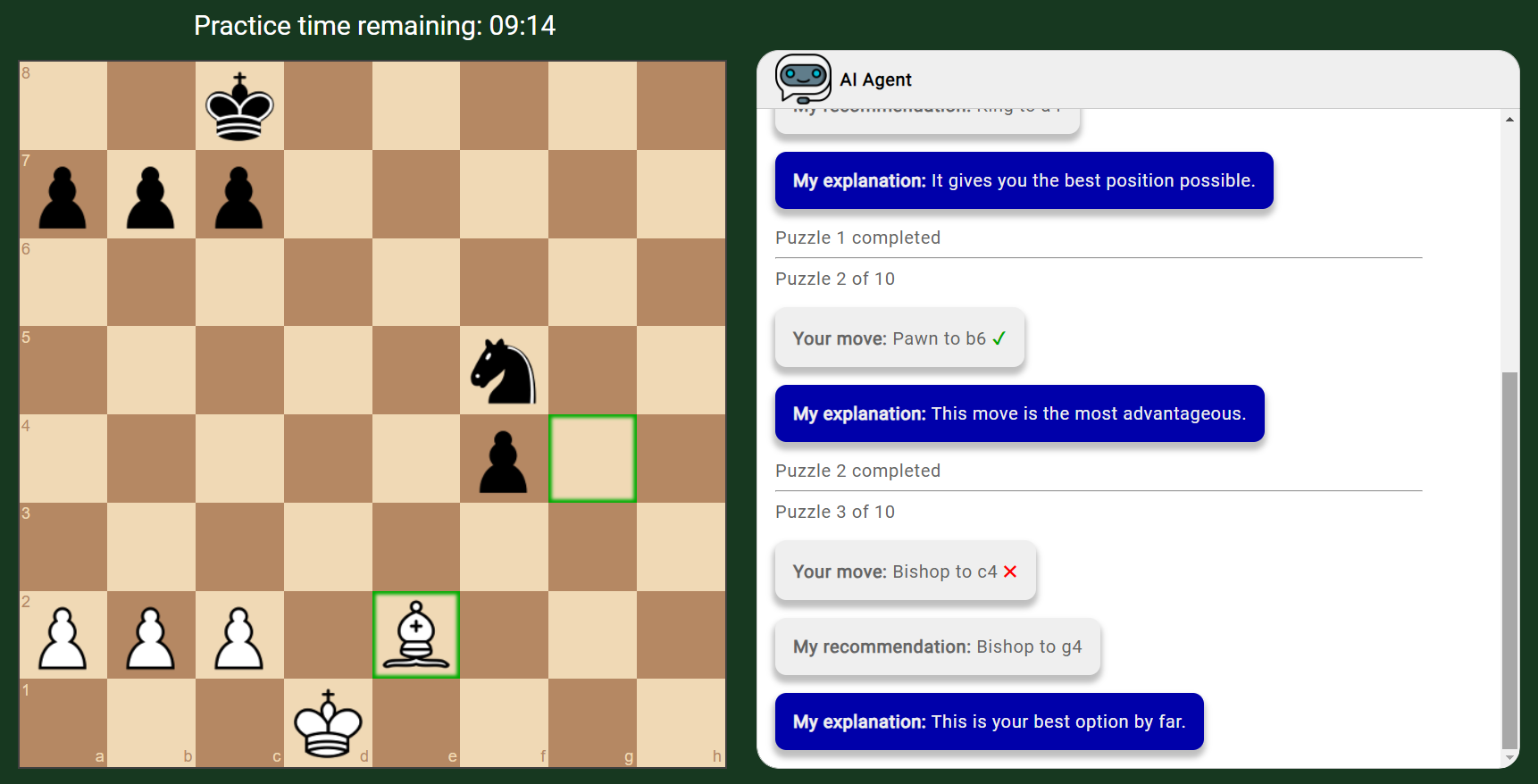}
        \caption{Practice section.}
        \label{fig:website_practice}
    \end{subfigure}
    \begin{subfigure}{\columnwidth}
        \centering
        \includegraphics[width=\textwidth]{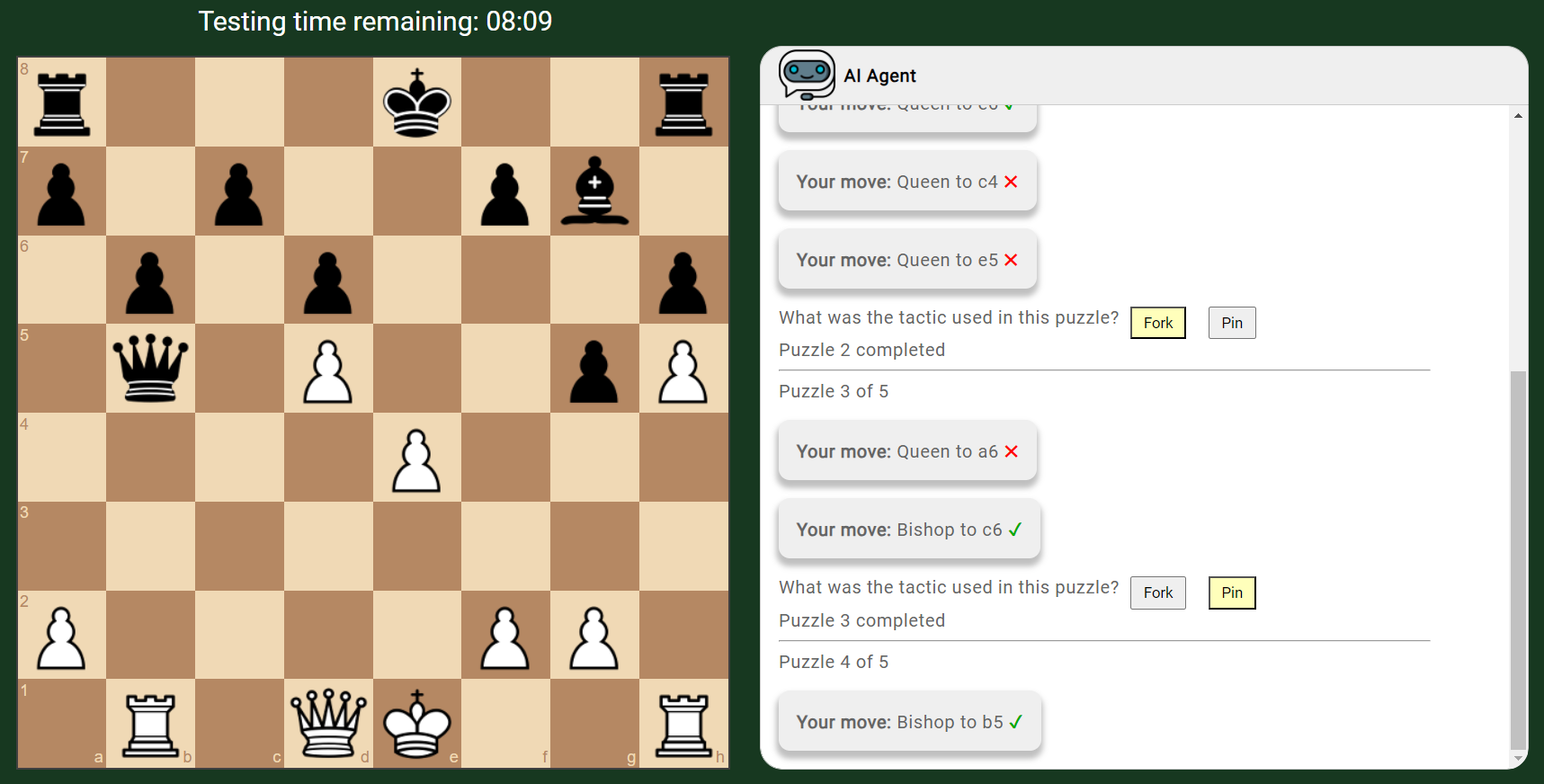}
        \caption{Testing section.}
        \label{fig:website_testing}
    \end{subfigure}
    \caption{Screenshots of the agent interface during the study.}
    \label{fig:website_sections}
\end{figure}

While solving these puzzles, users received recommendations and associated explanations via a dialog (chat) box to the right of the chessboard. The AI agent's messages in the chat box were separated by the puzzle number. Figure~\ref{fig:website_sections} shows the agent interface for the practice (training) and testing sections of chess puzzles. During both sections, if the user selected a piece, all valid moves for that piece were highlighted to assist the user in avoiding illegal move choices. During the practice section, incorrect moves were followed by move recommendations and, for two of the conditions, by associated explanations. An example of a placebic explanation used in the study would be, ``This move is the most advantageous,'' which is obvious to the user since they knew the computer model already recommended the best move. The actionable explanation for the same puzzle would be, ``This forks the king and rook with your pawn,'' which provides new information to the user beyond the fact that it was the best move. The explanations were shown in blue messages to grab the user's attention. The figure includes examples of placebic explanations. Correct and incorrect user choices were identified with ``green check'' or ``red cross'' marks, respectively.

\subsection{Experiment Design}
\label{sec:experiment-design}

\begin{figure}
    \centering
    \includegraphics[width=\textwidth]{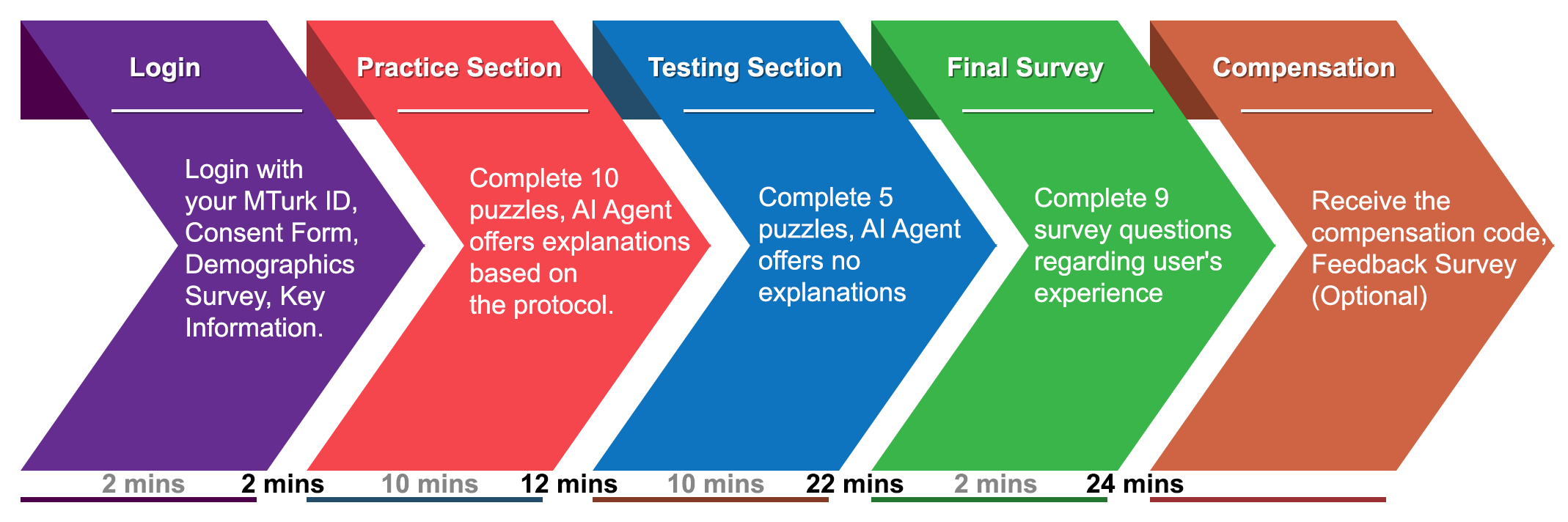}
    \caption{Experiment design.}
    \label{fig:experiment_design}
\end{figure}

This between-subjects study (the structure of which is given by Figure~\ref{fig:experiment_design}) was conducted using the Amazon Mechanical Turk (MTurk) platform to recruit online workers. MTurk is a popular hub for qualified workers to complete Human Intelligence Tasks (HITs) in exchange for money. The minimum requirements for worker participation in our study were a 95\% HIT approval rate, a minimum of 100 HITs completed, and residency in the United States. The first two requirements followed the common suggestions for MTurk studies~\cite{berinsky2012evaluating, chandler2014nonnaivete, hauser2016attentive, paolacci2010running, peer2014reputation}. The recruitment post also specified that users should be proficient in English and have a basic understanding of the rules of chess.

\begin{figure}
    \centering
    \begin{subfigure}{.49\columnwidth}
        \centering
        \includegraphics[width=\textwidth]{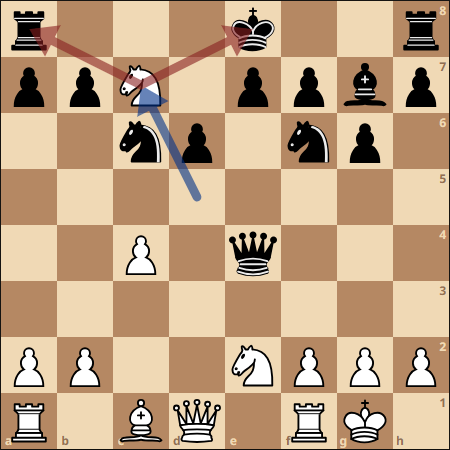}
        \caption{Fork.}
        \label{fig:example_fork}
    \end{subfigure}
    \begin{subfigure}{.49\columnwidth}
        \centering
        \includegraphics[width=\textwidth]{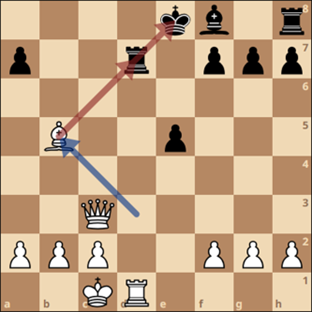}
        \caption{Pin.}
        \label{fig:example_pin}
    \end{subfigure}
    \caption{Examples of the tactics featured in the puzzles.}
    \label{fig:example_tactics}
\end{figure}

Once a user logged in using their unique MTurk ID, they read and agreed to the informed consent form, and then they completed a basic demographics survey asking them their age, gender, ethnicity, and highest level of education. They were also asked to describe their chess skills as either Beginner, Intermediate, or Expert. After the demographics survey, the workers were sent to a Key Information page containing important reminders about the AI agent and the two tactics they will see in the upcoming puzzles, given by Figure~\ref{fig:example_tactics}.

Each user engaged with up to 10 training puzzles and 5 testing puzzles for a total of 15 chess puzzles across 2 tactics: forks and pins. In the practice section, users predicted the best move and then received the model's best move and an explanation. After the practice section, the users demonstrated understanding in a test section by searching for the move that the model would rate as the best. 

Once they completed the testing section, they were presented with 9 statements to which they rated their agreement levels on a 7-point Likert scale~\cite{dawes2008data}. The subjects were then presented with their completion code to enter on the MTurk website to validate their participation. They also had the opportunity to provide optional feedback to the researchers.

We rewarded users with additional bonus compensation beyond the base pay rate. In the pilot study, this bonus was calculated by the number of test puzzles they completed with no errors. We found that this did not properly motivate users to perform carefully or thoughtfully once they had made a mistake. Subsequently, we changed the incentive structure so that users were rewarded for every test puzzle they completed minus the number of mistakes made within the puzzle, up to a maximum of 5 mistakes. Thus, each mistake subtracted 20\% from the test puzzle's maximum bonus. All test puzzles were weighted evenly. Puzzles not completed before the time limit were not added to the user's bonus compensation. Additionally, we provided more secure measures to ensure that users were reading the important information on each page, such as a 45-second timer on the key information page and highlighting the explanations to the groups in which they were given.

We are also currently in the process of improving the study design even more. Specifically, we are interested in testing out an additional performance metric in which users are asked for the tactic used in the testing puzzles they complete. This is featured in Figure~\ref{fig:website_testing}, though it was not included in the trials presented in this chapter.

\section{Results \& Data Analysis}
\label{sec:results-and-data-analysis}

\subsection{Demographics}
\label{sec:demographics}

\begin{figure}
    \centering
    \begin{subfigure}{.49\columnwidth}
        \centering
        \includegraphics[width=\textwidth]{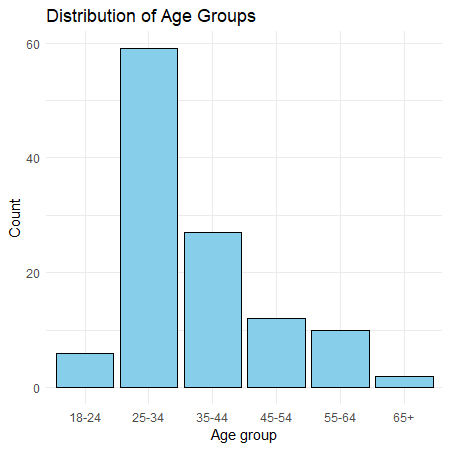}
        \caption{Age.}
        \label{fig:dist_age}
    \end{subfigure}
    \begin{subfigure}{.49\columnwidth}
        \centering
        \includegraphics[width=\textwidth]{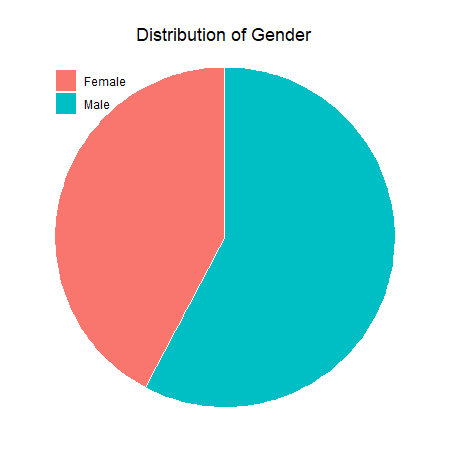}
        \caption{Gender.}
        \label{fig:dist_gender}
    \end{subfigure}
    \begin{subfigure}{.49\columnwidth}
        \centering
        \includegraphics[width=\textwidth]{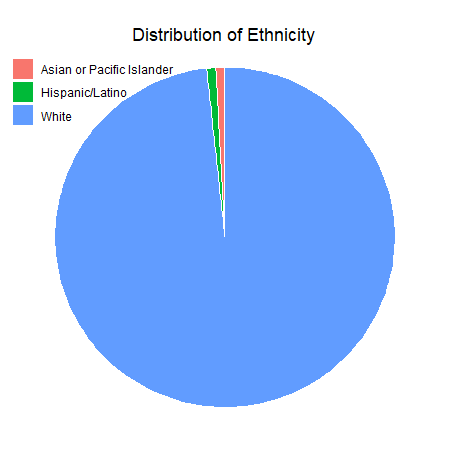}
        \caption{Ethnicity.}
        \label{fig:dist_ethnicity}
    \end{subfigure}
    \begin{subfigure}{.49\columnwidth}
        \centering
        \includegraphics[width=\textwidth]{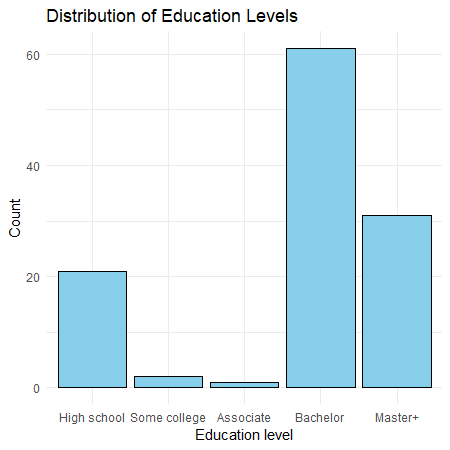}
        \caption{Education.}
        \label{fig:dist_education}
    \end{subfigure}
    \caption{Demographics data.}
    \label{fig:demographics}
\end{figure}

Immediately after signing into the study with their MTurk ID, our subjects took a brief demographics survey. The results from the objective survey questions are summarized in Figure~\ref{fig:demographics}.

All subjects were required to be at least 18 years old, and most of them were between the ages of 25 and 44. About 58\% of them identified as male while the remaining 42\% identified as female. The vast majority of the participants, or about 98\%, identified as white. Additionally, most of the subjects had either Bachelor's or graduate degrees.

\begin{figure}
    \centering
    \includegraphics[width=.6\textwidth]{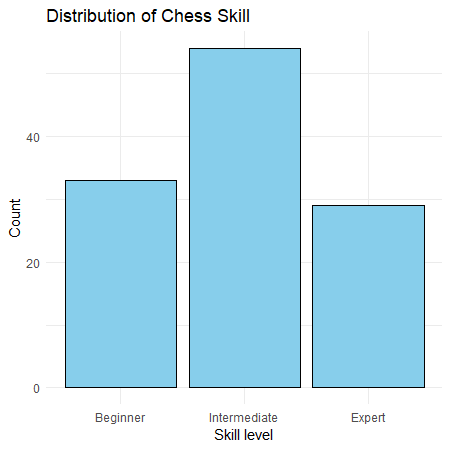}
    \caption{Self-identified skill level.}
    \label{fig:dist_skill}
\end{figure}

The last survey question asked the users to categorize their own skill level in chess, as the study's recruitment post stressed that participants must know the basic rules of chess. Using this subjective measure provided a nearly symmetric distribution among the three groups, with a peak at the intermediate level. This can be seen in Figure~\ref{fig:dist_skill}. However, we did not find any evidence that these self-perceived skill levels correlated with their performance.

\subsection{Performance}
\label{sec:performance}

The score of a single puzzle attempt was calculated as the inverse of the geometric mean of the number of user moves and the total duration of the attempt in seconds, given by Equation~\ref{eq:score}.
\begin{equation}
    \texttt{score} = \frac{1}{\sqrt{\texttt{num\_moves} \cdot \texttt{num\_seconds}}}
\label{eq:score}
\end{equation}
This ensured that a user scored the most for completing a puzzle with as few mistakes and as quickly as possible.

There was a total of 959 practice puzzle attempts by 108 distinct users and 371 testing puzzle attempts by 97 distinct users. We then filtered out the attempts which were not completed, took longer than 5 minutes to complete, or contained more than 10 mistakes by the user. This ensured that the remaining data constituted genuine user attempts. After filtering, there were 933 practice puzzle attempts among 102 distinct users and 208 testing puzzle attempts among 70 distinct users. Table~\ref{tab:counts_attempts_protocol}
\begin{table}
    \centering
    \begin{tabular}{ll|rrr}
        \hline
        \textit{Filtered?} & \textit{Section} & \textit{None} & \textit{Placebic} & \textit{Actionable} \\
        \hline
        No & Practice & 265 & 328 & 366 \\
        No & Testing & 111 & 125 & 135 \\
        Yes & Practice & 258 & 320 & 355 \\
        Yes & Testing & 63 & 70 & 75 \\
        \hline
    \end{tabular}
    \caption{Number of puzzles attempted within each protocol.}
    \label{tab:counts_attempts_protocol}
\end{table}
and Table~\ref{tab:counts_users_protocol}
\begin{table}
    \centering
    \begin{tabular}{ll|rrr}
        \hline
        \textit{Filtered?} & \textit{Section} & \textit{None} & \textit{Placebic} & \textit{Actionable} \\
        \hline
        No & Practice & 30 & 36 & 42 \\
        No & Testing & 29 & 31 & 37 \\
        Yes & Practice & 29 & 34 & 39 \\
        Yes & Testing & 20 & 25 & 25 \\
        \hline
    \end{tabular}
    \caption{Number of distinct users within each protocol.}
    \label{tab:counts_users_protocol}
\end{table}
break down these counts by protocol. We note that these totals are significantly larger than similar studies within the literature~\cite{eiband2019impact, langer1978mindlessness}.

\begin{figure}
    \centering
    \begin{subfigure}{.49\columnwidth}
        \centering
        \includegraphics[width=\textwidth]{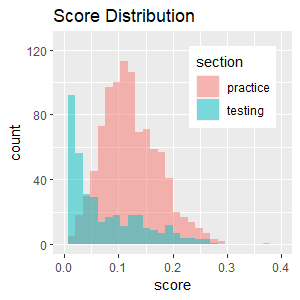}
        \caption{Unfiltered.}
        \label{fig:hist_score_unfiltered}
    \end{subfigure}
    \begin{subfigure}{.49\columnwidth}
        \centering
        \includegraphics[width=\textwidth]{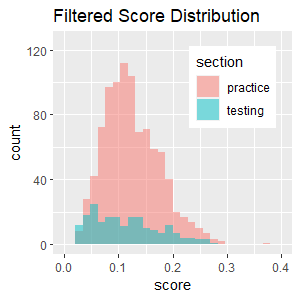}
        \caption{Filtered.}
        \label{fig:hist_score_filtered}
    \end{subfigure}
    \caption{Score distributions.}
    \label{fig:hist_score}
\end{figure}

As you can see from Figure~\ref{fig:hist_score_unfiltered},
the testing scores are highly skewed right from what appears to be an exponential distribution. This is due to a high number of incomplete or brute-forced puzzle attempts. In Figure~\ref{fig:hist_score_filtered}, the resulting distributions after filtering are significantly less skewed, and we believe they are more representative of thoughtful attempts.

\begin{figure}
    \centering
    \begin{subfigure}{.49\columnwidth}
        \centering
        \includegraphics[width=\textwidth]{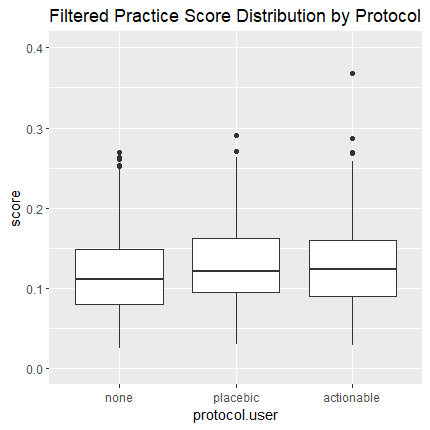}
        \caption{Practice.}
        \label{fig:boxplot_score_practice}
    \end{subfigure}
    \begin{subfigure}{.49\columnwidth}
        \centering
        \includegraphics[width=\textwidth]{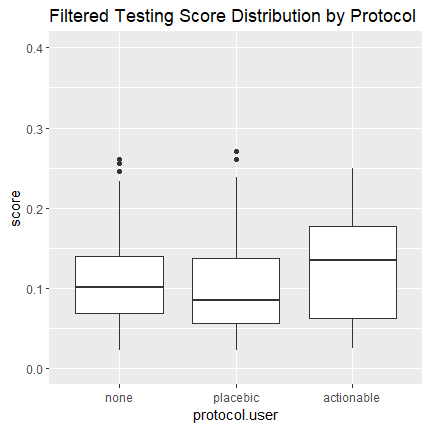}
        \caption{Testing.}
        \label{fig:boxplot_score_testing}
    \end{subfigure}
    \caption{Boxplots of scores grouped by protocol.}
    \label{fig:boxplot_score}
\end{figure}

The resulting score distributions, separated by protocol, can be seen in Figure~\ref{fig:boxplot_score}.
It appears that the {\em None} group performed worse in the practice section, while the {\em Actionable} group performed better in the testing section.

\begin{figure}
    \centering
    \begin{subfigure}{.49\columnwidth}
        \centering
        \includegraphics[width=\textwidth]{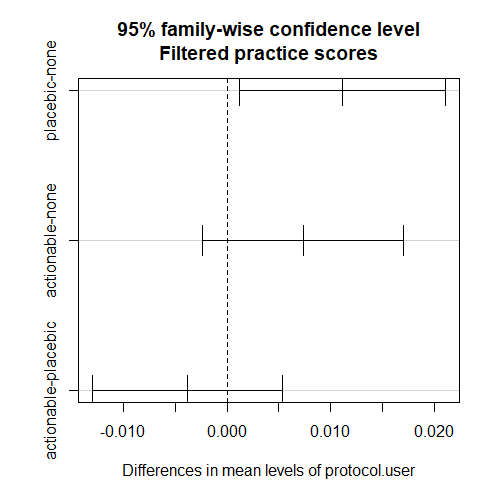}
        \caption{Practice.}
        \label{fig:tukey_score_practice}
    \end{subfigure}
    \begin{subfigure}{.49\columnwidth}
        \centering
        \includegraphics[width=\textwidth]{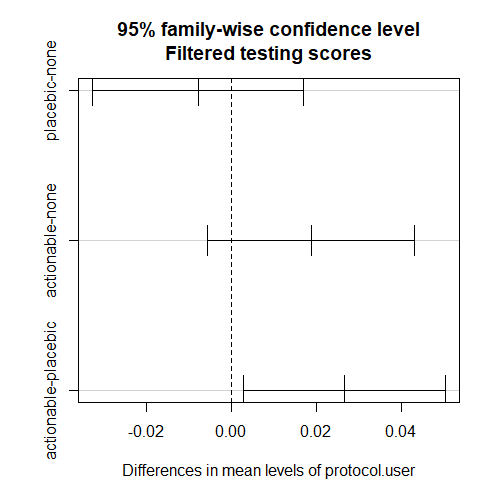}
        \caption{Testing.}
        \label{fig:tukey_score_testing}
    \end{subfigure}
    \caption{Tukey's HSD results for filtered scores.}
    \label{fig:tukey_score}
\end{figure}

Using analysis of variance (ANOVA), we find that there was a significant difference ($p = 0.0304$) in the mean scores of the different protocols in the practice section. According to Figure~\ref{fig:tukey_score_practice},
Tukey's honestly significant difference test (HSD) shows a significant difference ($p = 0.0238096$) between the {\em Placebic} and {\em None} groups.

We also find that there was a significant difference ($p = 0.026$) in the mean scores within the testing section. In Figure~\ref{fig:tukey_score_testing}, Tukey's HSD shows a significant difference ($p = 0.0235599$) between the {\em Actionable} and {\em Placebic} groups.

\begin{figure}
    \centering
    \includegraphics[width=.75\textwidth]{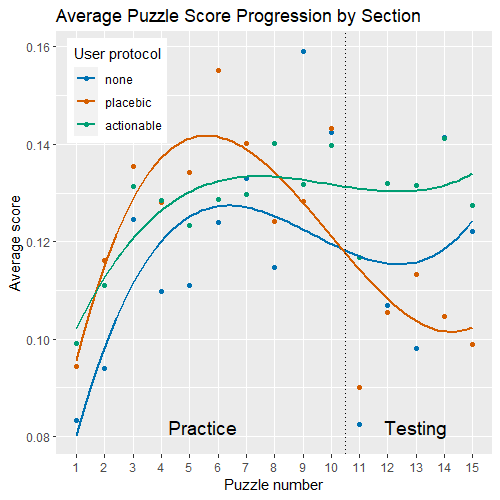}
    \caption{Average puzzle score progression across both sections of the study.}
    \label{fig:score_progression}
\end{figure}

It is also important to see if one or more the protocols realized any shift in user understanding as the study progressed. This piece of analysis serves to parse out the groups which may have consisted of, say, a significant number of under-performing individuals who still exhibited a larger increase in competence compared to other groups. Figure~\ref{fig:score_progression}
shows how the average score for each puzzle changed as a function of the puzzle number, with cubic-polynomial trendlines fitted to each of the three groups using a linear model. These third-order estimates are simple yet representative since scores in both study sections demonstrated an increase while the transition between the sections caused a significant decrease.

Scores with actionable explanations seemed to stay consistent after the transition from the practice section to testing, while the scores with placebic and no explanations dropped off significantly. This suggests that the actionable explanations were the most effective at providing sufficient knowledge for the users to perform well on their own.

These observations are confirmed by the numbers found in Table~\ref{tab:section_diffs},
\begin{table}
    \centering
    \begin{tabular}{lrrr}
        \hline
        \textit{Protocol} & \textit{Practice Score} & \textit{Testing Score} & \textit{\% Difference} \\
        \hline
        none & 0.119 & 0.110 & -7.3 \\
        placebic & 0.130 & 0.102 & -21.3 \\
        actionable & 0.126 & 0.129 & 2.2 \\
        \hline
    \end{tabular}
    \caption{Average performance differences between both sections.}
    \label{tab:section_diffs}
\end{table}
which show how users presented with actionable explanations saw a slight increase in their performance across the two sections while all others' scores decreased. In fact, placebic explanations created a massive drop-off in average score, illustrating their lack of substantive information and inability to engender learning in their human subjects.

\subsection{Survey Responses}
\label{sec:survey-responses}

The three survey variables were user satisfaction with the practice section, user satisfaction with the AI agent, and the explanatory power of the AI agent's explanations. The users were given three statements per variable, to which they selected their agreement level from a 7-point Likert scale. There were 29 respondents from the {\em None} group, 32 from the {\em Placebic} group, and 35 from the {\em Actionable} group.

\begin{figure}
    \centering
    \begin{subfigure}{.49\textwidth}
        \centering
        \includegraphics[width=\textwidth]{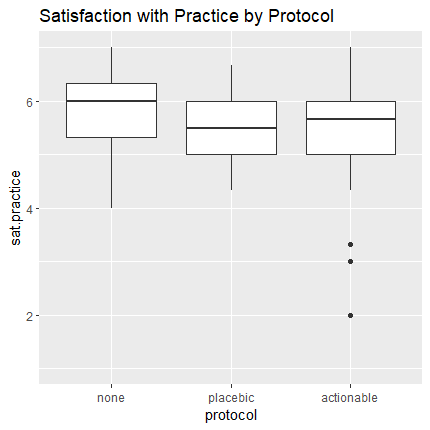}
        \caption{Satisfaction with practice.}
        \label{fig:boxplot_satPractice}
    \end{subfigure}
    \begin{subfigure}{.49\textwidth}
        \centering
        \includegraphics[width=\textwidth]{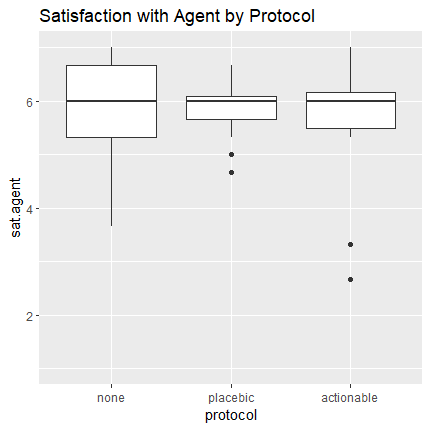}
        \caption{Satisfaction with the agent.}
        \label{fig:boxplot_satAgent}
    \end{subfigure}
    \begin{subfigure}{.49\textwidth}
        \centering
        \includegraphics[width=\textwidth]{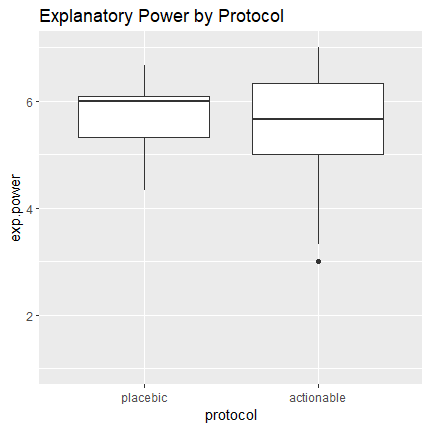}
        \caption{Explanatory power.}
        \label{fig:boxplot_expPower}
    \end{subfigure}
    \caption{Boxplots of survey variables grouped by protocol.}
    \label{fig:boxplot_surveyVariables}
\end{figure}

Figure~\ref{fig:boxplot_surveyVariables}
shows the distributions of each variable within each of the protocols. Upon an initial look, it appears that the {\em Placebic} and {\em Actionable} protocols might have a lower mean response than the {\em None} protocol. Additionally, the {\em Actionable} protocol appears to have less explanatory power in the eyes of the users.

\begin{figure}
    \centering
    \begin{subfigure}{.49\textwidth}
        \centering
        \includegraphics[width=\textwidth]{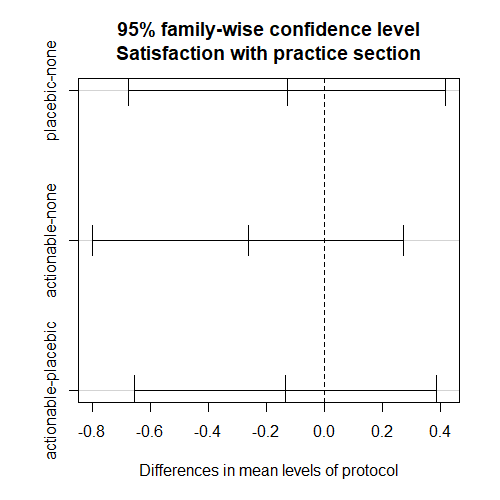}
        \caption{Satisfaction with practice.}
        \label{fig:tukey_satPractice}
    \end{subfigure}
    \begin{subfigure}{.49\textwidth}
        \centering
        \includegraphics[width=\textwidth]{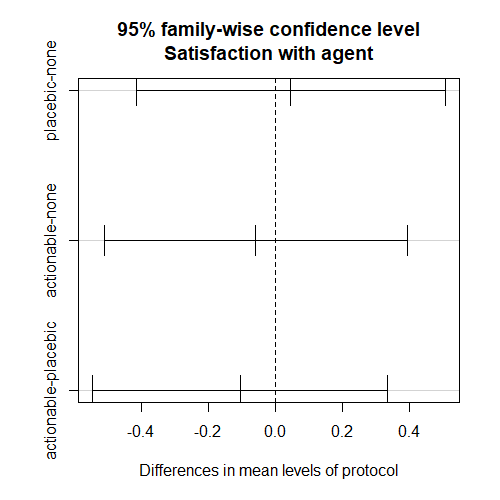}
        \caption{Satisfaction with the agent.}
        \label{fig:tukey_satAgent}
    \end{subfigure}
    \begin{subfigure}{.49\textwidth}
        \centering
        \includegraphics[width=\textwidth]{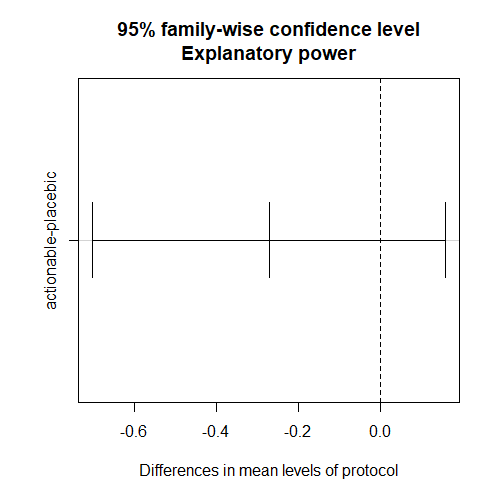}
        \caption{Explanatory power.}
        \label{fig:tukey_expPower}
    \end{subfigure}
    \caption{Tukey's HSD results on the survey variables.}
    \label{fig:tukey_surveyVariables}
\end{figure}

Using ANOVA once again, we find that there was not a significant difference ($p = 0.504$) in the mean responses for user satisfaction with the practice section. Likewise, there was not a significant difference in satisfaction with the agent ($p = 0.851$), nor was there a significant difference in explanatory power ($p = 0.213$). Figure~\ref{fig:tukey_surveyVariables}
shows all of these pairwise mean differences.

\section{Discussion}
\label{sec:discussion}

The results showed that there was not a significant difference between user satisfaction of placebic and actionable explanations, meaning that \textbf{hypothesis H2 is supported} by the data. We believe this demonstrates that users are equally satisfied with any number of coherent explanations written in their native language. So long as the explanations appear relevant to the task at hand, their quality is irrelevant to the human satisfaction of receiving them.

Additionally, the data \textbf{does support hypothesis H4}, since those receiving actionable explanations achieved significantly higher puzzle scores than those receiving placebic explanations, on average. We believe that this is due to the better inherent quality of the actionable explanations when compared to the placebic ones. Those in the {\em Actionable} group learned concrete ways to look for the two tactics in the testing puzzles. On the other hand, the {\em Placebic} subjects did not gain any useful knowledge from the practice section that could be applied to improve their scores in the testing section.

Furthermore, the users who received actionable explanations maintained and even improved their puzzle scores on average. This cannot be said for the participants from the other two groups. Those receiving no explanations at all performed understandably worse in the testing section, as they no longer received the agent's guidance. The placebic explanations seemed to give users confidence in the agent, and subsequently their own abilities, since they performed quite well in the practice section. However, these same individuals fell flat during the testing section, as they had gained no expertise in the matter to support the decisions they had to make alone. This goes to show that placebic explanations can potentially bring about more harm than good when placed in the wrong domain -- one where individual or human-agent team performance is critical.

The survey responses for user-perceived explanatory power showed no significant difference between the {\em Actionable} and {\em Placebic} group means. Therefore, hypothesis \textbf{H5 is not supported} by the results of this study. However, we believe this is still quite a valuable result of this study. The overall lack of expertise exhibited from the participants suggests that they were incapable of consciously knowing for themselves what was or was not a powerful explanation, while their performance was unconsciously improved or worsened because of it. This is further evidence showing the inadequacy of user surveys.

This study had some challenges demonstrating the differences we expected between the two explanation groups and the no-explanation group. Firstly, our results did not show that receiving placebic explanations was significantly more satisfying than receiving no explanations at all. Additionally, the test puzzle scores between the {\em Actionable} and {\em None} groups were not significantly different, though they were close to the significance threshold. Thus, \textbf{hypotheses H1 and H3 are not supported} by the data. We believe the absence of this expected difference stemmed from possible annoyance, or frustration, of reading the explanations. When the agent provided an explanation, all the elements on the screen except for the agent's most recent messages would blur temporarily. Thus, users assigned the {\em None} protocol could make the correct move more quickly and with less interruption than their counterparts. Perhaps a different study design where the user experience is more similar between groups would solve this problem.

\section{Conclusions}
\label{sec:conclusions}

This study successfully reached several of the most important conclusions it set out to solidify. Additionally, it discovered some contrary results which will prove vital for further research into the evaluation of computer-generated explanations.

The results of our objective score measure showed that the actionable explanations provided a significant boost to user performance. There were also a couple of score differences, though statistically insignificant, that leaned in the direction of the above conclusion, such as the mean differences between the {\em Actionable} and {\em None} groups in Figure~\ref{fig:tukey_score}.

Conversely, there were no significant differences in the subjective metrics gathered during the final survey questions. Even the insignificant differences leaned toward the opposite conclusions one would expect: that the {\em Actionable} agent may have less explanatory power than the {\em Placebic} agent and that the {\em None} agent may be the most satisfying.

With these two observations, we can say that a similar study that only uses user surveys to evaluate the quality of the model's explanations would be insufficient and might conclude that placebic explanations are just as good as actionable ones, and both of which are equally as effective as providing no explanations at all. This was the primary premise and motivation for our work and has been demonstrated through our experiments and associated results.

Using our earlier domain analysis, we can determine that a large number of users did not consciously find the domain important enough to warrant explanations. Alternatively, a highly skilled user would find the decision explanations obvious from the solution, rendering the actionable explanations equally as useful as the other two options. For the unskilled user, the agent's recommendation may be too complex to explain properly.

\section{Future Work}
\label{sec:future-work}

While explanations can serve to enhance various facets of user engagement in a domain, one of the key aspects of our domain is the impact of explanations to enhance user comprehension of generalizable domain knowledge. In particular, we use explanations to clarify recommended actions to teach concepts that the user can gainfully apply in other relevant scenarios. Hence, the main purpose of explanations here is to aid user learning and teach reusable knowledge. The performance metrics presented above capture one direct effect of such learning, i.e., improved problem-solving, In the future, related metrics can be designed to measure how conscious users are of the concepts learned. For example, we have started asking the user to identify the particular chess concept being applied, fork or pin, in each of the testing scenarios. We believe that such domain-specific metrics based on the concept being taught can be useful supplementary measures of the efficacy of explanation types.

We have used user satisfaction with the practice section and the agent, as well as the explanatory power of the explanations provided as survey factors. While we considered these factors to be of primary value for the current study, various other agent and process-related factors, such as comprehensibility, relevance, trust in the agent, etc. can be of interest. We plan to add survey questions on these factors in our future experiments.

The current study follows a between-subjects design, where each user receives either no explanations, placebic explanations, or actionable explanations. It can be instructive to evaluate a within-subjects study design to see how the same user responds to different explanation types. However, ordering effects of explanation types have to be accounted for to cover the effects of either learning through increased exposure to a concept or user fatigue over prolonged experimentation sessions.

With our earlier domain analysis in mind, we are currently in the process of designing and experimenting with two types of domains that can be used to compare different explanation types: (a) teaching concepts, which is similar to the chess domain, e.g., a virtual escape room or a social security scenario optimization task; and, (b) domains where explanations are key to improving human-agent interactions, e.g., explaining agent actions in human-agent teams, heterogeneous search-and-rescue teams, etc.

\section*{Acknowledgements}
We thank the The University of Tulsa Cyber Fellows program for partially funding this work. We also thank Selim Karaoglu for his assistance with this study.

\bibliographystyle{unsrt}
\bibliography{references}

\begin{thebibliography}{10}

\bibitem{aggarwal2022has}
Karan Aggarwal, Maad~M Mijwil, Abdel-Hameed Al-Mistarehi, Safwan Alomari, Murat G{\"o}k, Anas M~Zein Alaabdin, Safaa~H Abdulrhman, et~al.
\newblock Has the future started? the current growth of artificial intelligence, machine learning, and deep learning.
\newblock {\em Iraqi Journal for Computer Science and Mathematics}, 3(1):115--123, 2022.

\bibitem{langer2021we}
Markus Langer, Daniel Oster, Timo Speith, Holger Hermanns, Lena K{\"a}stner, Eva Schmidt, Andreas Sesing, and Kevin Baum.
\newblock What do we want from explainable artificial intelligence (xai)?--a stakeholder perspective on xai and a conceptual model guiding interdisciplinary xai research.
\newblock {\em Artificial Intelligence}, 296:103473, 2021.

\bibitem{renftle2022explaining}
Moritz Renftle, Holger Trittenbach, Michael Poznic, and Reinhard Heil.
\newblock Explaining any ml model?--on goals and capabilities of xai.
\newblock {\em arXiv preprint arXiv:2206.13888}, 2022.

\bibitem{rosenfeld2019explainability}
Avi Rosenfeld and Ariella Richardson.
\newblock Explainability in human--agent systems.
\newblock {\em Autonomous Agents and Multi-Agent Systems}, 33:673--705, 2019.

\bibitem{lavender2023relative}
Bryan Lavender, Sami Abuhaimed, and Sandip Sen.
\newblock Relative effects of positive and negative explanations on satisfaction and performance in human-agent teams.
\newblock In {\em The International FLAIRS Conference Proceedings}, volume~36, 2023.

\bibitem{miller2019explanation}
Tim Miller.
\newblock Explanation in artificial intelligence: Insights from the social sciences.
\newblock {\em Artificial intelligence}, 267:1--38, 2019.

\bibitem{mohseni2021quantitative}
Sina Mohseni, Jeremy~E Block, and Eric Ragan.
\newblock Quantitative evaluation of machine learning explanations: A human-grounded benchmark.
\newblock In {\em 26th International Conference on Intelligent User Interfaces}, pages 22--31, 2021.

\bibitem{mohseni2021multidisciplinary}
Sina Mohseni, Niloofar Zarei, and Eric~D Ragan.
\newblock A multidisciplinary survey and framework for design and evaluation of explainable ai systems.
\newblock {\em ACM Transactions on Interactive Intelligent Systems (TiiS)}, 11(3-4):1--45, 2021.

\bibitem{chromik2020taxonomy}
Michael Chromik and Martin Schuessler.
\newblock A taxonomy for human subject evaluation of black-box explanations in xai.
\newblock {\em Exss-atec@ iui}, 1, 2020.

\bibitem{keane2021if}
Mark~T Keane, Eoin~M Kenny, Eoin Delaney, and Barry Smyth.
\newblock If only we had better counterfactual explanations: Five key deficits to rectify in the evaluation of counterfactual xai techniques.
\newblock {\em arXiv preprint arXiv:2103.01035}, 2021.

\bibitem{belaid2022we}
Mohamed~Karim Belaid, Eyke H{\"u}llermeier, Maximilian Rabus, and Ralf Krestel.
\newblock Do we need another explainable ai method? toward unifying post-hoc xai evaluation methods into an interactive and multi-dimensional benchmark.
\newblock {\em arXiv preprint arXiv:2207.14160}, 2022.

\bibitem{rosenfeld2021better}
Avi Rosenfeld.
\newblock Better metrics for evaluating explainable artificial intelligence.
\newblock In {\em Proceedings of the 20th international conference on autonomous agents and multiagent systems}, pages 45--50, 2021.

\bibitem{mohseni2018human}
Sina Mohseni, Jeremy~E Block, and Eric~D Ragan.
\newblock A human-grounded evaluation benchmark for local explanations of machine learning.
\newblock {\em arXiv preprint arXiv:1801.05075}, 2018.

\bibitem{eiband2019impact}
Malin Eiband, Daniel Buschek, Alexander Kremer, and Heinrich Hussmann.
\newblock The impact of placebic explanations on trust in intelligent systems.
\newblock In {\em Extended abstracts of the 2019 CHI conference on human factors in computing systems}, pages 1--6, 2019.

\bibitem{ehsan2021explainability}
Upol Ehsan and Mark~O Riedl.
\newblock Explainability pitfalls: Beyond dark patterns in explainable ai.
\newblock {\em arXiv preprint arXiv:2109.12480}, 2021.

\bibitem{nourani2019effects}
Mahsan Nourani, Samia Kabir, Sina Mohseni, and Eric~D Ragan.
\newblock The effects of meaningful and meaningless explanations on trust and perceived system accuracy in intelligent systems.
\newblock In {\em Proceedings of the AAAI Conference on Human Computation and Crowdsourcing}, volume~7, pages 97--105, 2019.

\bibitem{berinsky2012evaluating}
Adam~J Berinsky, Gregory~A Huber, and Gabriel~S Lenz.
\newblock Evaluating online labor markets for experimental research: Amazon. com's mechanical turk.
\newblock {\em Political analysis}, 20(3):351--368, 2012.

\bibitem{chandler2014nonnaivete}
Jesse Chandler, Pam Mueller, and Gabriele Paolacci.
\newblock Nonna{\"\i}vet{\'e} among amazon mechanical turk workers: Consequences and solutions for behavioral researchers.
\newblock {\em Behavior research methods}, 46:112--130, 2014.

\bibitem{hauser2016attentive}
David~J Hauser and Norbert Schwarz.
\newblock Attentive turkers: Mturk participants perform better on online attention checks than do subject pool participants.
\newblock {\em Behavior research methods}, 48:400--407, 2016.

\bibitem{paolacci2010running}
Gabriele Paolacci, Jesse Chandler, and Panagiotis~G Ipeirotis.
\newblock Running experiments on amazon mechanical turk.
\newblock {\em Judgment and Decision making}, 5(5):411--419, 2010.

\bibitem{peer2014reputation}
Eyal Peer, Joachim Vosgerau, and Alessandro Acquisti.
\newblock Reputation as a sufficient condition for data quality on amazon mechanical turk.
\newblock {\em Behavior research methods}, 46:1023--1031, 2014.

\bibitem{dawes2008data}
John Dawes.
\newblock Do data characteristics change according to the number of scale points used? an experiment using 5-point, 7-point and 10-point scales.
\newblock {\em International journal of market research}, 50(1):61--104, 2008.

\bibitem{langer1978mindlessness}
Ellen~J Langer, Arthur Blank, and Benzion Chanowitz.
\newblock The mindlessness of ostensibly thoughtful action: The role of ``placebic'' information in interpersonal interaction.
\newblock {\em Journal of personality and social psychology}, 36(6):635, 1978.

\end{thebibliography}

\end{document}